\documentclass[aps,prb,twocolumn,noeprint]{revtex4-1}
\usepackage{graphicx}
\usepackage{amsmath}
\usepackage{tikz}
\PassOptionsToPackage{caption=false}{subfig}
\usepackage{subfig}
\usepackage{hyperref}
\hypersetup{colorlinks=true, citecolor=blue, urlcolor=blue, linkcolor=blue}
\DeclareMathOperator{\tr}{Tr}

\makeatletter
\renewcommand{\maketag@@@}[1]{\hbox{\m@th\normalsize\normalfont#1}}%
\makeatother

\newcommand{\bra}[1]{\ensuremath{\langle{#1}|\,}}
\newcommand{\ket}[1]{\ensuremath{\,|{#1}\rangle}}

\begin{document}
\title{Non-renewal statistics in quantum transport through the eyes of first-passage and waiting time distributions}
\author{Samuel L. Rudge}
\author{Daniel S. Kosov}
\address{College of Science and Engineering, James Cook University, Townsville, QLD, 4811, Australia}

\begin{abstract}
The waiting time distribution has, in recent years, proven to be a useful statistical tool for characterising transport in nanoscale quantum transport. In particular, as opposed to moments of the distribution of transferred charge, which have historically been calculated in the long-time limit, waiting times are able to detect non-renewal behaviour in mesoscopic systems. They have failed, however, to correctly incorporate backtunneling events. Recently, a method has been developed that can describe unidirectional and bidirectional transport on an equal footing: the distribution of first-passage times. Rather than the time between successive electron tunnelings, the first-passage refers to the first time the number of extra electrons in the drain reaches $+1$. Here, we demonstrate the differences between first-passage time statistics and waiting time statistics in transport scenarios where the waiting time either cannot correctly reproduce the higher order current cumulants or cannot be calculated at all. To this end, we examine electron transport through a molecule coupled to two macroscopic metal electrodes. We model the molecule with strong electron-electron and electron-phonon interactions in three regimes: (i) sequential tunneling and cotunneling for a finite bias voltage through the Anderson model, (ii) sequential tunneling with no temperature gradient and a bias voltage through the Holstein model, and (iii) sequential tunneling at zero bias voltage and a temperature gradient through the Holstein model. We show that, for each transport scenario, backtunneling events play a significant role; consequently, the waiting time statistics do not correctly predict the renewal and non-renewal behaviour, whereas the first-passage time distribution does.
\end{abstract}

\maketitle
 
\section{Introduction}
Molecular electronic devices generally consist of some quantum system, such as a molecule or quantum dot, sandwiched between two macroscopic metal electrodes \cite{scheer2010molecular,Xiang2016}. One of the main objectives of mesoscopic physics is to use the particle transport through such devices to understand the nature of these quantum systems. Directly measuring transitions internal to the quantum system destroys any quantum coherence it may exhibit; instead, one must reconstruct the system behaviour from transport properties available from measurements of bath variables, such as the stationary particle current $\langle I \rangle$. Molecular devices are small enough, however, that thermal effects and stochastic single electron transport cause fluctuations in $\langle I \rangle$. Such fluctuations hold a dual interest in mesoscopic electron transport. On the one hand they impede the performance of electronic components that require reliable currents; but on the other hand they are largely determined by the molecule's underlying dynamics and can thus be used to probe quantum behaviour, which is difficult to observe. Consequently, multiple theoretical tools describing fluctutations in quantum electron transport have been developed \cite{Bagrets2003,Brandes2008,Saito2016,Budini2014,Esposito2010}. 

The full counting statistics (FCS) calculates cumulants $\langle\langle I^{k}\rangle\rangle$ of the distribution of total transferred charge $n$ in a fixed time interval. \cite{Levitov1996,Bagrets2003,Hassler2008,Nazarov2003} The second current cumulant, the zero-frequency noise, can reveal the internal dynamics of the molecule \cite{Blanter2000,Mahe2010,Bednorz2010}. For example, super-Poissonian shot noise has many origins: it can arise from telegraphic switching due to spin-polarised leads \cite{Ptaszynski2017,Ptaszynski2017a} or inelastic cotunneling \cite{Thielmann2005a,Kaasbjerg2015}; negative differential resistance due to asymmetric coupling \cite{Thielmann2005}; the dynamical channel blockade \cite{Belzig2005}; or avalanching electrons due to interactions with a vibrational mode \cite{Koch2005}. Higher order cumulants are necessary to fully characterise the transport when the current distribution is non-Gaussian \cite{Bruderer2014,Reulet2003,Bomze2005}. The major limitation of fixed time tools like the FCS is that they have historically been calculated in the infinite time, or zero-frequency, limit \cite{Budini2014}. Although finite frequency, and thus finite time, FCS have been developed \cite{Emary2007,Ubbelohde2012,Galaktionov2003,Nagaev2004,Pilgram2004,Galaktionov2011,Marcos2010,Salo2006}, it is numerically and analytically easier to work in the zero-frequency regime. In recent years, however, statistical tools that operate in fluctuating time but fixed $n$ have arisen, motivated by the long time limitations of FCS.

The most prominent flucutating time tool is the waiting time distribution (WTD) $w(\tau)$: the conditional probability density that, given an electron has tunneled to the drain electrode, the next electron will tunnel to the drain after a time interval $\tau$ \cite{Brandes2008}. Since the inception of waiting times in mesoscopic electron transport, a common question has been whether the WTD contain information distinct from the FCS. It has been shown that the large deviation rate functions describing zero-frequency FCS are equivalent to those for waiting times \cite{Budini2014}; the WTD cumulants $\langle\langle\tau\rangle\rangle$ can be combined to exactly reproduce the zero-frequency current cumulants. \cite{Budini2011,Esposito2010,Albert2011} These relations only hold, however, when the renewal assumption is satisfied: $w(\tau_{1},\tau_{2})=w(\tau_{1})w(\tau_{2})$, where $w(\tau_{1},\tau_{2})$ is the joint probability for successive waiting times. When subsequent waiting times are correlated the renewal assumption is violated and these relationships no longer hold; the WTD contains information separate to the FCS. Interestingly, even in Markovian systems non-renewal behaviour can arise from telegraphic switching \cite{Ptaszynski2017,Ptaszynski2017a}, superconductivity \cite{Albert2016}, quantum coherence \cite{Dasenbrook2015}, and the presence of strong interactions \cite{Rudge2018,Kosov2017b}. WTD measurements are experimentally accessible via sophisticated real-time single-electron detection techniques \cite{Gustavsson2006,Gustavsson2009,Gustavsson2008,Zbydniewska2015,Lu2003}; or alternatively the WTD may be reconstructed from low-order current correlation functions \cite{Haack2015}.

The WTD in quantum electron transport can be calculated via scattering theory \cite{Albert2011,Albert2012,Chevallier2016,Albert2014,Albert2016,Haack2014,Dasenbrook2015} and non-equilibrium Green's functions \cite{Tang2014,SeoaneSouto2015}. In order to connect it with other fluctuating time statistics, however, we focus on WTDs calculated from quantum master equations, which rely on quantum jump operators defined from the Liouvillian \cite{Brandes2008,Rudge2018,Rudge2016,Rudge2016a,Kosov2016,Thomas2013,Sothmann2014,Kosov2017b,Ptaszynski2017a,Rajabi2013,Walldorf2018,Potanina2017,Welack2009,Ptaszynski2017,Kosov2018a}. Unfortunately, WTDs calculated in this manner are so far only able to reproduce $\langle\langle I^{k}\rangle\rangle$ ($k>1$), as according to renewal theory, for unidirectional transport. For bidirectional transport, one may define WTDs between tunnelings to and from the drain; the total average current, regardless of whether the transport is renewal, is then $\langle I \rangle = \frac{1}{\langle\tau\rangle_{F}} - \frac{1}{\langle\tau\rangle_{B}}$, where $\langle\tau\rangle_{F}$ is the average waiting time between forward tunnelings, conditioned upon the initial probability of a tunneling to the drain, and $\langle\tau\rangle_{B}$ is the corresponding backward tunneling average waiting time. It is not clear, however, how to obtain the higher order total current cumulants from these two WTDs. In some cases, furthermore, the processes that contribute to the current do not appear in the Liouvillian; for example, elastic cotunneling transfers electrons from the source to the drain in the same quantum process without altering the state of the intermediate system \cite{Braggio2006,Kaasbjerg2015,Utsumi2006,Emary2009}. To fully describe the transport in these cases, one must use the $n$-resolved master equation \cite{Gurvitz1998}; but waiting times calculated from this method \cite{Thomas2013,Potanina2017,SeoaneSouto2015} are only defined for unidirectional transport \cite{Rudge2018}. Bidirectional transitions play an important role in mesoscopic electron transport \cite{Fujisawa2006}, so it is desirable to have a fluctuating time statistic applicable to this regime.

Consider instead the first-passage time distribution (FPTD): the distribution of times until the $\text{\it{net}}$ number of transferred electrons in the drain first reaches some number $n$. First-passage times have recently gained traction in non-equilibrium thermodynamics; first-passage time theory has been developed for fluctuations in entropic variables describing stationary Markovian processes \cite{Bauer2014,Roldan2015,Neri2017,Garrahan2017,Saito2016,Ptaszynski2018}. This work, and subsequent experimental and theoretical research from Singh et al. \cite{Singh2018}, has predicted and verified fluctutation relations for first-passage times. Saito and Dhar \cite{Saito2016} have developed first-passage time notation for Markovian quantum systems, which is suitable for mesoscopic electron transport described by $n$-resolved master equations. Ptaszynski \cite{Ptaszynski2018} has also recently brought first-passage times into line with FCS and WTDs, by developing a method to calculate correlations between subsequent first-passage times; deriving exact relationships between the FPTD cumulants and the FCS cumulants; and demonstrating that these relationships disappear when the renewal assumption is violated. The expectation now is that first-passage times are able to analyse the non-renewal behaviour of molecular systems similarly to how WTDs have been used, except that first-passage times are applicable to bidirectional transport. Considering these developments, it is interesting to compare the WTD and the FPTD in systems and regimes where one expects bidirectional transitions. 

This paper's purpose is to demonstrate the usefulness of FPTDs by comparing them to the WTD in regimes where the WTD either cannot be calculated or does not correctly recreate the FCS. We demonstrate this in three transport scenarios. First, cotunneling through an Anderson impurity is considered. Several authors have theoretically \cite{Thielmann2005,Kaasbjerg2015} and experimentally \cite{Okazaki2013,Zhang2007,Onac2005} shown that for a particular voltage range super-Poissonian noise arises from telegraphic switching induced by inelastic cotunneling processes. In this voltage range backtunneling processes can occur; so until now it has remained impenetrable to waiting times. One would expect that telegraphic switching would be accompanied by positive correlations in the subsequent waiting times, as the transport switches between channels with different characteristic $\langle\tau\rangle$; however, we demonstrate that the transport in this regime is actually characterised by small negative correlations in the first-passage times. Second, we examine sequential tunneling through the Holstein model due to a voltage bias. Waiting times have already been theoretically calculated for this scenario \cite{Kosov2017b,Kosov2016}; we demonstrate that those calculations do not correctly reproduce the FCS and that the correlations between the first-passage times are slightly smaller than those between the waiting times. Finally, we consider sequential tunneling through the Holstein model with zero bias voltage but with a temperature gradient: a scenario in which backtunneling events majorly contribute to the transport \cite{De2018}.

The paper is organised as follows. Section \ref{Statistical tools} outlines the three relevant statistical tools: FCS, the WTD, and the FPTD. Section \ref{Methods} describes the dynamics of the three transport scenarios, written in the form of Markovian master equations. Section \ref{Results} contains results from the statistical tools calculated for the three transport scenarios, while Section \ref{Conclusion} summarises the work and the Appendix details cotunneling calculations and vibration jump operators.

Throughout the manuscript we use natural units: $\hbar=k_{B}=e=1$.


\section{Statistical tools} \label{Statistical tools}

\subsection{$n$-resolved master equation} \label{n resolved master equation subsection methods}

We consider a general quantum system weakly coupled to a number of thermal reservoirs, which is described by a Markovian master equation for the reduced system density matrix:

\begin{align}
\dot{\mathbf{P}}(t) & = \mathbf{L}\mathbf{P}(t), \label{General master equation}
\end{align}
where $\{\mathbf{P}\}_{l}(t)$ is the probability that the system is in discrete state $l$ and $\mathbf{L}$ is the Liouvillian containing the system dynamics. Throughout the paper, we refer to this as the $\it{\text{standard}}$ master equation, to distinguish it from the $n$-resolved master equation. We ignore coherence between system states, so that Eq.\eqref{General master equation} reduces to a rate equation written in the system basis. The system moves from state $l$ to state $k$ via tunneling of particles or quasi-particles from or to the thermal reservoirs, and the transition rate for this process is $\Gamma_{kl}$. The Liouvillian off-diagonals are $[\mathbf{L}]_{kl} = \Gamma_{kl}$ and the diagonals are $[\mathbf{L}]_{kk} = -\sum\limits_{l\neq k}\Gamma_{lk}$. 

The system starts in some state $\mathbf{P}(0)$, so that the general solution of Eq.\eqref{General master equation} is

\begin{align}
\mathbf{P}(t) = e^{\mathbf{L}t}\mathbf{P}(0). \label{General master equation solution}
\end{align}
We assume that the system has a unique stationary state $\bar{\mathbf{P}}$, which is the null vector satisfying $\mathbf{L}\bar{\mathbf{P}} = 0$.

In order to calculate the FCS, FPTD, and WTD, it is necessary to resolve the master equation on the net number of particles $n$ transferred to the drain electrode:

\begin{align}
\dot{\mathbf{P}}(n,t) & = \sum_{n'}\mathbf{L}(n-n')\mathbf{P}(n,t), \label{General n-resolved master equation}
\end{align}
where $\{\mathbf{P}(n,t)\}_{k}$ now represents the probability that, at time $t$, the net number of extra electrons in the drain is $n$ and the system is in state $k$. It is important to note that despite the term ``extra electrons'' $n$ may be zero, a negative integer, or a positive integer. From here, following Ptaszynski's notation \cite{Ptaszynski2018}, we will refer to $n$ as the jump number. Sequential and cotunneling processes add or remove at most one electron to or from the drain, so $n-n' = 0,\pm 1$. The Liouvillian can then be split into two jump operators containing all the transitions that move electrons forward to the drain $\mathbf{J}_{F}$ and all the transitions that move electrons back from the drain $\mathbf{J}_{B}$, and a part containing the remaining dynamics $\mathbf{L}_{0} = \mathbf{L} - \mathbf{J}_{F} - \mathbf{J}_{B}$. The $n$-resolved master equation is then

\begin{align}
\dot{\mathbf{P}}(n,t) & = \mathbf{L}_{0}\mathbf{P}(n,t) + \mathbf{J}_{F}\mathbf{P}(n-1,t) + \mathbf{J}_{B}\mathbf{P}(n+1,t). \label{General n-resolved master equation expanded}
\end{align} 
The Fourier transform of $\mathbf{P}(n,t)$ 

\begin{align}
\mathbf{P}(\chi,t) & = \sum_{n}e^{in\chi}\mathbf{P}(n,t), \text{and conversely}  \label{Fourier transformed probability vector} \\
\mathbf{P}(n,t) & = \frac{1}{2\pi}\int_{0}^{2\pi} d\chi e^{-in\chi} \mathbf{P}(\chi,t), \label{Inverse Fourier transformed probability vector}
\end{align}
conveniently introduces a counting field $\chi$ that transforms Eq.\eqref{General n-resolved master equation} from an infinite set of coupled differential equations to a solvable problem \cite{Bagrets2003,Nazarov1999}:

\begin{align}
\dot{\mathbf{P}}(\chi,t) & = \mathbf{L}(\chi)\mathbf{P}(\chi,t), \text{ where }  \label{Chi dependent master equation} \\
\mathbf{L}(\chi) & = \mathbf{L}_{0} + \mathbf{J}_{F}e^{i\chi} + \mathbf{J}_{B}e^{-i\chi}. \label{Full Liouvillian Chi}
\end{align}

Now, $\mathbf{P}(\chi,t)$ time-evolves according to the solution of Eq.\eqref{Chi dependent master equation}, with the initial condition $\mathbf{P}(\chi,0) = \bar{\mathbf{P}}$ since it is assumed that counting begins at $t=0$ when the system has already reached the steady state:

\begin{align}
\mathbf{P}(\chi,t) & = e^{\mathbf{L}(\chi)t}\bar{\mathbf{P}}. \label{Chi dependent master equation solution}
\end{align}

\subsection{Full counting statistics} \label{FCS subsection methods}

The generating function for the cumulants of transferred charge is 

\begin{align}
M(\chi,t) & = \ln \sum_{n} P(n,t) e^{in\chi}, \label{Cumulant generating function}
\end{align} 
where $P(n,t) = \tr[\mathbf{P}(n,t)]$ is the probability that $n$ extra electrons are transferred to the drain in the time interval $[0,t]$. Throughout the paper we use the superoperator notation $\tr[\mathbf{A}] = \left(\mathbf{I},\mathbf{A}\right)$, where $\mathbf{A}$ is a column vector of length $m$ and $\mathbf{I} = (1,1,1,...)$ is a row vector of length $m$. Consequently, Eq.\eqref{Cumulant generating function} transforms to 

\begin{align}
M(\chi,t) & = \ln \sum_{n} \left(\mathbf{I},\mathbf{P}(n,t)\right) e^{in\chi}, 
\end{align} 
with the inverse Fourier transform 

\begin{align}
\left(\mathbf{I},\mathbf{P}(n,t)\right) & = \frac{1}{2\pi}\int_{0}^{2\pi} d\chi e^{-in\chi} e^{M(\chi,t)}.
\end{align} 

Examining the trace of Eq.\eqref{Inverse Fourier transformed probability vector}, one sees that the part corresponding to the cumulant generating function is 

\begin{align}
M(\chi,t) & = \ln \left(\mathbf{I},e^{\mathbf{L}(\chi)t}\bar{\mathbf{P}}\right). \label{Final form cumulant generating function}
\end{align} 

Theoretically, one can now calculate all current cumulants in any time interval by direct differentiation of $M(\chi,t)$ in the limit $\chi\rightarrow0$: $\langle\langle I^{k} \rangle\rangle = \frac{1}{t}(-i)^{k}\frac{\partial^{k}}{\partial\chi^{k}}M(\chi,t)\Big|_{\chi=0}$, where $t$ is the measurement time. The second cumulant, for example, is $\langle I^{2}\rangle - \langle I\rangle^{2}$. It is well known that, except for the simplest systems, these calculations remain too difficult at finite $t$. Bagrets and Nazarov have shown that in the long-time limit, the cumulant generating function reduces to 

\begin{align}
M(\chi,t) & = t\Lambda_{max},
\end{align}
where $\Lambda_{max}$ is the eigenvalue of $\mathbf{L}(\chi)$ with the largest real part  \cite{Bagrets2003,Flindt2008}.

We focus on the Fano factor, the second cumulant scaled by the average current, which measures the zero-frequency noise power $\mathcal{S}(0)$ compared to that of a Poissonian process:

\begin{align}
F & = \frac{\langle\langle I^{2} \rangle\rangle}{\langle I \rangle} \label{Fano factor definition} \\
& = \frac{\mathcal{S}(0)}{2\langle I \rangle}.
\end{align}

For Poissonian processes $F=1$; sub-Poissonian and super-Poissonian processes are characterised by $F<1$ and $F>1$, respectively. 

\subsection{Waiting time distribution} \label{WTD subsection methods}

If instead one wishes to examine the distribution of waiting times between subsequent tunnelings to the drain, then one may use the formula developed by Brandes \cite{Brandes2008}:

\begin{align}
w_{F}(\tau) & = \frac{\left(\mathbf{I},\mathbf{J}_{F}e^{(\mathbf{L}-\mathbf{J}_{F}) \tau}\mathbf{J}_{F}\bar{\mathbf{P}}\right)}{\left(\mathbf{I},\mathbf{J}_{F}\bar{\mathbf{P}}\right)}, \label{Brandes WTD}
\end{align}
The waiting time distribution between tunnelings $\text{\it{from}}$ the drain, $w_{B}(\tau)$, is similarly defined. 

There are tunneling events that do not appear in the standard master equation, yet change the drain number $n$. To include these events in the WTD, one must use the $n$-resolved master equation and the definition of the WTD from the idle time probability \cite{Thomas2013,Rudge2018,Albert2012,VanKampen1981}:

\begin{align}
w_{F}(\tau) & = \frac{1}{p}\frac{\partial^{2}}{\partial\tau^{2}}\Pi(\tau), \label{WTD definition ITP}
\end{align}
where $p$ is the probability for observing a tunneling to the drain and $\Pi(\tau)$ is the idle time probability: the probability that no tunneling events to the drain are observed by time $\tau$. 

In the case of unidirectional transport, when $\mathbf{J}_{B} = 0$ and all backtunneling transitions are safely neglected from the Liouvillian, $p = \langle I \rangle$ and the idle time probability is obtained from $n$-resolved master equation:

\begin{align}
\Pi(\tau) & = \lim_{i\chi\rightarrow -\infty} \left(\mathbf{I},e^{\mathbf{L}(\chi)\tau}\bar{\mathbf{P}}\right). \label{Idle time probability}
\end{align}
Now, $\mathbf{L}(\chi) = \mathbf{L}_{0} + \mathbf{J}_{F}e^{i\chi}$ and no backtunneling transition rates are included in $\mathbf{L}_{0}$. Evidently, backtunneling events will be accompanied by an $e^{-i\chi}$ factor that will diverge in the limit $i\chi \rightarrow -\infty$. Of course, if the transport is unidirectional in the backward direction as opposed to the forward direction, then the same definition may be used with the limit $i\chi \rightarrow \infty$; either way only one direction of tunneling events may be included. Using Eq.\eqref{Idle time probability} in the waiting time definition from Eq.\eqref{WTD definition ITP} yields the same form of the WTD as developed by Brandes, except now the exponent $\mathbf{L}(0) - \mathbf{J}_{F}$ is obtained from the $n$-resolved master equation; when the transport is unidirectional the two methods coincide. The only difference between them is that when the jump operators can be defined in terms of a standard master equation Brandes' method can calculate both $w_{F}(\tau)\text{ and }w_{B}(\tau)$, whereas Eq.\eqref{WTD definition ITP} cannot.

The Laplace transform of the WTD is

\begin{align}
\tilde{w}(z) & =  \frac{\left(\mathbf{I},\mathbf{J}(z - \mathbf{L}_{0})^{-1}\mathbf{J}\bar{\mathbf{P}}\right)}{\left(\mathbf{I},\mathbf{J}\bar{\mathbf{P}}\right)}, 
\end{align}
from which the cumulants are calculated:
\begin{align}
\langle\langle\tau^{k}\rangle\rangle & = (-1)^{k} \frac{d^{k}}{dz^{k}}\ln\tilde{w}(z)\Big|_{z=0}.
\end{align}
We have included the notation $\mathbf{L}_{0} = \mathbf{L} - \mathbf{J}$ as shorthand for the exponent in Eq.\eqref{Brandes WTD}. 

The average current can be reconstructed from the average waiting times:

\begin{align}
\langle I \rangle = \frac{1}{\langle\tau\rangle_{F}} - \frac{1}{\langle\tau\rangle_{B}}, \label{WTD Current definition}
\end{align}
where $\langle\tau\rangle_{F}\text{ and }\langle\tau\rangle_{B}$ are the average forward and backward waiting times, respectively. Waiting times calculated from an $n$-resolved only satisfy Eq.\eqref{WTD Current definition} for unidirectional transport\cite{Rudge2018}. The equality in Eq.\eqref{WTD Current definition} holds regardless of whether the transport is renewal, which follows from examining the forward and backward currents separately. The forward current, reconstructed from the WTD, is

\begin{align}
\langle I \rangle_{F} & = \sum_{k=1}^{\infty} \frac{k}{\langle\tau_{1}+\tau_{2}+\hdots+\tau_{k}\rangle_{F}}P(k) \label{WTD expanded current definition} \\
\end{align}
where $P(k)$ is the probability that the forward current is $\frac{k}{\langle\tau_{1}+\tau_{2}+\hdots+\tau_{k}\rangle_{F}}$. The average waiting times are defined via the joint probability distribution:

\begin{widetext}
\begin{align}
\langle\tau_{1}+\tau_{2}+\hdots+\tau_{k}\rangle & = \int d\tau_{k} \hdots \int d\tau_{2} \int d\tau_{1} \left(\tau_{1} + \tau_{2} + \hdots \tau_{k}\right) w(\tau_{1},\tau_{2},\hdots,\tau_{k}) \\
& = \sum_{p=1}^{k} \int d\tau_{p} \: \tau_{p} w(\tau_{p}) \label{WTD Current defintion shortening} \\
& = k\langle\tau\rangle,
\end{align}
\end{widetext}
where we have made no assumption on the renewal nature of the transport. Eq.\eqref{WTD expanded current definition} then reduces to 
\begin{align}
\langle I \rangle_{F} & = \frac{1}{\langle\tau\rangle_{F}},
\end{align}
assuming that $\{P(k)\}$ spans the entire current space. The same reasoning follows for the backward current. 

The higher order current cumulants, in comparison, are only reconstructable from the WTD for unidirectional transport; for example,  the randomness parameter 

\begin{align}
R & = \frac{\langle\langle\tau^{2}\rangle\rangle}{\langle\tau\rangle^{2}}, \label{RP WTD}
\end{align}

is equal to the Fano factor defined in Eq.\eqref{Fano factor definition} for unidirectional transort satisfying the renewal assumption. The Pearson correlation coefficient measures the strength of correlations between successive waiting times:

\begin{align}
p & = \frac{\langle\tau_{1}\tau_{2}\rangle - \langle\tau\rangle^{2}}{\langle\langle\tau^{2}\rangle\rangle}. \label{pearson WTD definition}
\end{align}

When $p$ is nonzero, the renewal assumption is violated and the equality between Eq.\eqref{RP WTD} and Eq.\eqref{Fano factor definition} no longer applies. 

The first moment of the joint distribution $\langle\tau_{1}\tau_{2}\rangle$ is calculated similarly to the cumulants:
\begin{align}
\langle\tau_{1}\tau_{2}\rangle & = \frac{\partial}{\partial z_{1}}\frac{\partial}{\partial z_{2}}\tilde{w}(z_{1},z_{2})\Big|_{z_{1}=z_{2}=0} \\
& =  \frac{\left(\mathbf{I},\mathbf{J}\mathbf{L}_{0}^{-2}\mathbf{J}\mathbf{L}_{0}^{-2}\mathbf{J}\bar{\mathbf{P}}\right)}{\left(\mathbf{I},\mathbf{J}\bar{\mathbf{P}}\right)}.
\end{align}

\subsection{First-passage time distribution} \label{FPTD subsection methods}

Unlike the WTD, which calculates the time interval between successive electron tunnelings to the drain, the FPTD calculates the time interval between a tunneling to the drain and when the jump number next reaches a specified integer $n$. Necessarily, the FPTD includes backward tunneling events, and it is defined from the $n$-resolved master equation. To derive the FPTD we follow the method neatly encapsulated by Saito and Dhar in the framework of Markovian master equations \cite{Saito2016}, and explicitly explained by Ptaszynski \cite{Ptaszynski2018}.

We return to the $\chi$-dependent Liouvillian defined in Eq.\eqref{Full Liouvillian Chi} and start the time-evolution in some arbitrary state $\mathbf{P}(0)$. The $n$-resolved probability vector is written using Eq.\eqref{Chi dependent master equation} and Eq.\eqref{Chi dependent master equation solution}:

\begin{align}
\mathbf{P}(n,t) & = \frac{1}{2\pi}\int_{0}^{2\pi} d\chi e^{-in\chi} e^{\mathbf{L}(\chi)t}\mathbf{P}(0) \\
& = \mathbf{T}(n|t)\mathbf{P}(0). \label{Rate equation FPTD}
\end{align}

Now, $\mathbf{T}(n|t) = \frac{1}{2\pi}\int_{0}^{2\pi} d\chi e^{-in\chi} e^{\mathbf{L}(\chi)t}$ is the evolution operator for $\mathbf{P}(n,t)$, which in Laplace space is 

\begin{align}
\tilde{\mathbf{T}}(n|z) & = \frac{1}{2\pi}\int_{0}^{2\pi} d\chi e^{-in\chi} \left[z - \mathbf{L}\right]^{-1}. \label{Transition matrix laplace definition}
\end{align}
The matrix element $\left[\mathbf{T}(n,t)\right]_{kl}$ is the conditional probability that, given the system is initially in state $l$ at time $t=0$, it will be in state $k$ at time $t$ and the jump number will have reached $n$ for the first time. Saito and Dhar \cite{Saito2016}, Ptaszynski\cite{Ptaszynski2018}, and Singh et al.\cite{Singh2018} have all found exact expressions for $\tilde{\mathbf{T}}(n|z)$ for simple single reset systems, in which a tunneling to the drain leaves the molecule empty. In general, though, evaluating Eq.\eqref{Transition matrix laplace definition} requires a numerical contour integration, which is the approach we take. 

Alternatively, $\mathbf{P}(n,t)$ is obtainable by considering the column vector of first-passage time probabilities $\mathbf{F}(n|\tau)$. Here, $\left[\mathbf{F}(n|\tau)\right]_{k}$ is the probability density that at time $\tau$ the system is in state $k$ and $n$ extra electrons have been recorded for the first time in the drain. Premultiplication by the evolution operator $\mathbf{T}(0,t-\tau)\mathbf{F}(n,\tau)$ transitions the system in the interval $[\tau,t]$ to a final probability vector, during which no total change to $n$ is made. This contributes to the probability vector $\mathbf{P}(n,t)$; the final step integrates over all possible first-passage times:

\begin{align}
\mathbf{P}(n,t) & = \int^{t}_{0} d\tau \mathbf{T}(0,t-\tau)\mathbf{F}(n,\tau). \label{Volterra equation}
\end{align}

From here it is easier to work in Laplace space. The RHS of Eq.\eqref{Volterra equation} is a convolution, and using Eq.\eqref{Rate equation FPTD} one obtains 

\begin{align}
\tilde{\mathbf{T}}(n|z)\mathbf{P}(0) & = \tilde{\mathbf{T}}(0,z)\tilde{\mathbf{F}}(n,z).
\end{align}

In Laplace space, the total first-passage time probability is 

\begin{align}
\tilde{F}(n|z) & = \left(\mathbf{I},\tilde{\mathbf{T}}(0,z)^{-1}\tilde{\mathbf{T}}(n|z)\mathbf{P}(0)\right) \label{General FPTD Laplace}.
\end{align}

As Ptaszynski \cite{Ptaszynski2018} notes, in order to connect the FPTD to the WTD, the initial state $\mathbf{P}(0)$ needs to be the normalised system state just after a forward jump to the drain has occurred:

\begin{align}
\mathbf{P}(0) & = \frac{\mathbf{J}_{F}\bar{\mathbf{P}}}{\left(\mathbf{I},\mathbf{J}_{F}\bar{\mathbf{P}}\right)}.
\end{align}
Combining this choice of $\mathbf{P}(0)$ with Eq.\eqref{General FPTD Laplace}, we obtain 

\begin{align}
\tilde{F}(n|z) & = \frac{\left(\mathbf{I},\tilde{\mathbf{T}}(0,z)^{-1}\tilde{\mathbf{T}}(n|z)\mathbf{J}_{F}\bar{\mathbf{P}}\right)}{\left(\mathbf{I},\mathbf{J}_{F}\bar{\mathbf{P}}\right)} \label{FPTD Laplace Final},
\end{align}
where $F(n|\tau)$, defined as the inverse Laplace transform of Eq.\eqref{FPTD Laplace Final} $\mathcal{L}^{-1}\left\{\tilde{F}(n|z)\right\}$, is the probability that, given an initial tunneling to the drain occurred at some arbitrary time in the stationary state, the jump number first reached $n$ after a time interval $\tau$. For unidirectional transport, $F(1|\tau)$ exactly reproduces the WTD defined in Eq.\eqref{Brandes WTD}.

The FPTD in time-space is obtained from the numerical inverse Laplace transform of Eq.\eqref{FPTD Laplace Final}:

\begin{align}
F(n|\tau) & = \frac{1}{2\pi i}\lim_{R\rightarrow\infty}\int_{c-iR}^{c+iR}dz \: e^{z\tau}\frac{\left(\mathbf{I},\tilde{\mathbf{T}}(0,z)^{-1}\tilde{\mathbf{T}}(n|z)\mathbf{J}_{F}\bar{\mathbf{P}}\right)}{\left(\mathbf{I},\mathbf{J}_{F}\bar{\mathbf{P}}\right)}. \label{FPTD Time}
\end{align}

We focus, however, on the cumulants of the FPTD, which are again calculated from $\tilde{F}(n|z)$:

\begin{align}
\langle\langle\tau^{k}_{n}\rangle\rangle^{*} & = (-1)^{k} \lim_{z\rightarrow 0^{+}}\left[\frac{d^{k}}{dz^{k}}\ln\tilde{F}(n|z)\right], \label{CGF FPTD}
\end{align}
where the limit $z\rightarrow 0^{+}$ is required as $\mathbf{L}(\chi)$ is singular for $\chi = \{0,2\pi\}$ \cite{Ptaszynski2018}. We have introduced the $^{*}$ notation for FPTD cumulants to distinguish them from WTD cumulants, and used Ptaszynski's notation of $\langle\langle\tau^{k}_{n}\rangle\rangle^{*}$ referring to the $\text{k}^{th}$ cumulant of $F(n|\tau)$.

Ptaszynski has recently shown that when the transport is renewal the FPTD cumulants are linearly related:
\begin{align}
\langle\langle\tau^{k}_{n}\rangle\rangle^{*} & = n\langle\langle\tau^{k}_{1}\rangle\rangle^{*},
\end{align}
as well as deriving, similarly to the WTD, renewal relations between the FCS and the first-passage time cumulants \cite{Ptaszynski2018}. The average current is then 

\begin{align}
\langle I \rangle & = \frac{n}{\langle\tau_{n}\rangle^{*}} \\
& = \frac{1}{\langle\tau_{1}\rangle^{*}}. \label{FPTD Current definition}
\end{align}
Unlike the WTD, where the equality in Eq.\eqref{WTD Current definition} is satisfied regardless of renewal behaviour, the equality in Eq.\eqref{FPTD Current definition} is only true for renewal processes; the logic applied in Eq.\eqref{WTD Current defintion shortening} is not available to first-passage times since, for non-renewal processes, $F(2|\tau) \neq F(1|\tau;1|\tau')$.

One can also define a randomness parameter $R^{*}_{n}$ for each distribution $F(n|\tau)$:

\begin{align}
R^{*}_{n} & = n\frac{\langle\langle\tau_{n}^{2}\rangle\rangle^{*}}{\left(\langle\tau_{n}\rangle^{*}\right)^{2}}. \label{RP FPTD definition}
\end{align}
For renewal systems they all reduce to $R^{*}_{1}$, which is then equal to the Fano factor. Unlike $R$, we expect that the renewal relationships will hold in bidirectional transport. The Pearson correlation coefficient between subsequent first-passage times can be calculated using $F(2|\tau)$ \cite{Ptaszynski2018}:

\begin{align}
p^{*} & = 2\frac{R^{*}_{2}}{R^{*}_{1}} - 1. \label{pearson FPTD definition}
\end{align}

Throughout the majority of the paper we will not use the subscript $n$ for the cumulants, with the implication being that we are referring to cumulants of $F(1|\tau)$ unless explicitly stated.


\section{Model systems} \label{Methods}

In this section we introduce the three transport scenarios under consideration and define the rate equations describing their dynamics. Quantum rates are calculated using the T-matrix method, which is a suitable replacement for more rigorous approaches when the transport is Markovian; that is, when $T_{\alpha},V_{SD} \gg \gamma$, where $\gamma$ is the broadening of the molecular energy levels, $T_{\alpha}$ is the temperature of electrode $\alpha$, and $V_{SD}$ is the source-drain voltage bias \cite{Thielmann2005}. In order to use rate equations as opposed to comprehensive master equations \cite{Gurvitz1996,Gurvitz1996a,Gurvitz1998,Li2005,Harbola2006,Nazarov1993,Davies1993} we also operate under the secular approximation; coherence between molecular basis states are ignored. All scenarios describe the transport of electrons through a molecule weakly coupled to two macroscopic metal electrodes. The total Hamiltonian is 

\begin{align}
H = H_{S}+H_{D}+H_{M}+H_{T},
\end{align}
where $H_{M}$ is the Hamiltonian of the nanoscale quantum system; $H_{S}\text{ and }H_{D}$ are the source and drain Hamiltonians
\begin{align}
H_{S}+H_{D} = \sum_{\alpha = S,D}\sum_{k,\sigma}\varepsilon^{}_{\alpha,k,\sigma}a^{\dagger}_{\alpha,k,\sigma}a^{}_{\alpha,k,\sigma};
\end{align}
and $H_{T}$ is the system-electrode interaction
\begin{align}
\label{V}
H_{T} = \sum_{\alpha = S,D}\sum_{k, \sigma}t^{}_{\alpha,k,\sigma}(a^{\dagger}_{\alpha,k,\sigma}a^{}_{\sigma}+a_{\sigma}^{\dagger}a^{}_{\alpha,k,\sigma}).
\end{align}
The operator $a_{\alpha,k,\sigma}^{\dagger}$ creates an electron in electrode $\alpha$ in the single-particle state $k$ with spin $\sigma = \left\{\uparrow,\downarrow\right\}$ and free energy $\varepsilon_{\alpha,k,\sigma}$, while $a_{\alpha,k,\sigma}$ annihilates an electron with spin $\sigma$ from the same state in electrode $\alpha$, and $t_{\alpha,k,\sigma}$ are the tunneling amplitudes between the quantum system and the electrodes.

The molecule is described by the Anderson-Holstein model; an energy level, which can be spin split, is coupled to a vibrational mode such that electrons occupying the molecule experience electron-electron and electron-phonon interactions.

\begin{align}
H_{M} & = \sum_{\sigma}\varepsilon_{\sigma} a_{\sigma}^{\dagger}a_{\sigma}+U n_{\uparrow}n_{\downarrow} + \lambda(b^{\dagger} + b)n+\omega b^{\dagger}b. \label{Anderson-Holstein Hamiltonian}
\end{align}

The operators $a^{\dagger}_{\sigma}\text{ and }a_{\sigma}$ create and annihilate an electron in the single-particle state $\sigma$ with free energy $\varepsilon_{\sigma}$, while $b^{\dagger}\text{ and }b$ create and annihilate phonons with frequency $\omega$. The particle number operators are $n_{\sigma} = a^{\dagger}_{\sigma}a_{\sigma}$ and $n = \sum\limits_{\sigma} a^{\dagger}_{\sigma}a_{\sigma}$. The $\lambda(b^{\dagger} + b)n$ term describes the electron-phonon interaction, where $\lambda$ is the coupling strength. Likewise, $U$ is the Coulomb repulsion strength. 

The transition rate from the electrode-system many-body state $i$ to many-body state $f$ is 

\begin{align}
\Gamma_{fi} = 2\pi \big{|}\bra{f}T\ket{i}\big{|}^2\delta(E_{i}-E_{f}), \label{Generalised Fermi's golden rule}
\end{align} where the T-matrix is 
\begin{align}
T & = H_{T}+H_{T}\frac{1}{E_{i}-H_{0}+i\eta}H_{T} \nonumber \\
	& + H_{T}\frac{1}{E_{i}-H_{0}+i\eta}H_{T}\frac{1}{E_{i}-H_{0}+i\eta}H_{T}+... . \label{T matrix}
\end{align}

The leading order approximation is Fermi's golden rule, and describes sequential tunneling:
\begin{align}
\Gamma_{fi} = 2\pi \big{|}\bra{f}H_{T}\ket{i}\big{|}^2\delta(E_{i}-E_{f}).
\end{align}
Fermi's generalised golden rule \cite{Bruus2002,Timm2008} details the higher order terms; for example, cotunneling events are described by going to fourth order in $H_{T}$. 

\subsection{Anderson impurity}

For a rigid molecule with no electron-phonon coupling $\lambda = 0$, Eq.\eqref{Anderson-Holstein Hamiltonian} reduces to the Anderson model:

\begin{align}
H_{M} & = \sum_{\sigma}\varepsilon_{\sigma} a^{\dagger}_{\sigma}a_{\sigma} + U n_{\uparrow}n_{\bar{\downarrow}}. 
\end{align}

The molecular orbital is then in one of four system states; it can be empty $|0\rangle$, occupied by a single electron of either spin $|\sigma\rangle$, or occupied by two electrons of opposite spin $|2\rangle$, with probability vector $\mathbf{P}(\chi,t) = \left[P_{0}(\chi,t),P_{\uparrow}(\chi,t),P_{\downarrow}(\chi,t),P_{2}(\chi,t)\right]^{T}$. Including cotunneling and sequential tunneling events, the Liouvillian minus the forward and backward jump operators is 
\begin{widetext}
\begin{align}
\mathbf{L}_{0} & =  \left[\begin{array}{cccc}
 -(\Gamma_{\uparrow0}+\Gamma_{\downarrow0} +\Gamma_{00}^{(2)}) 
 & \Gamma_{0\uparrow}^{S} & \Gamma_{0\downarrow}^{S} & 0 \\
\\
\Gamma_{\uparrow 0}^{S} & -(\Gamma_{0\uparrow}+\Gamma_{2\uparrow} + \Gamma_{\uparrow\uparrow}^{(2)}+\Gamma_{\downarrow\uparrow}^{(2)}) & \Gamma_{\uparrow\downarrow}^{SS}+\Gamma_{\uparrow\downarrow}^{DD} & \Gamma_{\uparrow2}^{S}\\
\\
\Gamma_{\downarrow0}^{S} & \Gamma_{\downarrow\uparrow}^{SS}+\Gamma_{\downarrow\uparrow}^{DD} &  -(\Gamma_{0\downarrow}+\Gamma_{2\downarrow} + \Gamma_{\downarrow\downarrow}^{(2)} +  \Gamma_{\uparrow\downarrow}^{(2)}) & \Gamma_{\downarrow2}^{S}\\
\\
0 & \Gamma_{2\uparrow}^{S} & \Gamma_{2\downarrow}^{S} &  -(\Gamma_{\uparrow2}+\Gamma_{\downarrow2} +\Gamma_{22}^{(2)})
\end{array}\right], \nonumber \\
\end{align}
\end{widetext}
with the forward and backward jump operators defined as 

\begin{widetext}
\begin{align}
\mathbf{J}_{F} & = \left[\begin{array}{cccc}
\Gamma_{00}^{SD} & \Gamma_{0\uparrow}^{D} & \Gamma_{0\downarrow}^{D} & 0\\
& \\
0 & \Gamma_{\uparrow\uparrow}^{SD} & \Gamma_{\uparrow\downarrow}^{SD} & \Gamma_{\uparrow2}^{D}\\
& \\
0 & \Gamma_{\downarrow\uparrow}^{SD} & \Gamma_{\downarrow\downarrow}^{SD} & \Gamma_{\downarrow2}^{D}\\
& \\
0 & 0 & 0 & \Gamma_{22}^{SD}
\end{array}\right] \quad\quad \text{ and } & \mathbf{J}_{B} = \left[\begin{array}{cccc}
\Gamma_{00}^{DS} & 0 & 0 & 0\\
& \\
\Gamma_{\uparrow0}^{D} & \Gamma_{\uparrow\uparrow}^{DS} & \Gamma_{\uparrow\downarrow}^{DS} & 0\\
& \\
\Gamma_{\downarrow0}^{D} & \Gamma_{\downarrow\uparrow}^{DS} & \Gamma_{\downarrow\downarrow}^{DS} & 0\\
& \\
0 & \Gamma_{2\uparrow}^{D} & \Gamma_{2\downarrow}^{D} & \Gamma_{22}^{DS}
\end{array}\right].
\end{align}
\end{widetext}
The total sequential rates are 
\begin{align}
\Gamma_{kl} & = \Gamma_{kl}^{S} + \Gamma_{kl}^{D};
\end{align}
the total elastic cotunneling rates are 
\begin{align}
\Gamma_{ll}^{(2)} & = \Gamma_{ll}^{SD} + \Gamma_{ll}^{DS};
\end{align}
and the total inelastic cotunneling rates are 
\begin{align}
\Gamma_{kl}^{(2)} & = \Gamma_{kl}^{SD} + \Gamma_{kl}^{DS} + \Gamma_{kl}^{SS} + \Gamma_{lk}^{DD}.
\end{align}

The sequential rates are

\begin{align} 
\Gamma_{\sigma0}^{\alpha} & = \gamma^{\alpha} \: n_{F}(\varepsilon_{\sigma}-\mu_{\alpha}), \\
\Gamma_{0\sigma}^{\alpha} & = \gamma^{\alpha} \: \big{(}1-n_{F}(\varepsilon_{\sigma}-\mu_{\alpha})\big{)}, \\
\Gamma_{\sigma2}^{\alpha} & = \gamma^{\alpha} \: \big{(}1-n_{F}(\varepsilon_{\sigma}+U-\mu_{\alpha})\big{)}, & \text{and} \\
\Gamma_{2\sigma}^{\alpha} & = \gamma^{\alpha} \: n_{F}(\varepsilon_{\sigma}+U-\mu_{\alpha}), \\
\nonumber\end{align}
where $\gamma^{\alpha}=2\pi\big{|}t_{\alpha}\big{|}^2\rho(\varepsilon_{\alpha})$; $\rho(\varepsilon_{\alpha})$ is the density of states for the source and drain electrodes, which is assumed to be constant; and $n_{F}(\varepsilon-\mu_{\alpha})$ is the Fermi-Dirac distribution:
\begin{align}
n_{F}(\varepsilon-\mu_{\alpha}) &  = \frac{1}{1+e^{(\varepsilon-\mu_{\alpha})/T_{\alpha}}}.
\end{align}

During cotunneling, the electron undergoes two tunneling events in the same quantum process through an intermediate ``virtual'' energy level in the molecule. Elastic cotunneling leaves the molecule in the same energetic state as before the process, while inelastic cotunneling leaves the molecular orbital occupied by the same number of electrons but in an excited or de-excited state. The notation $\Gamma_{kl}^{\alpha\beta}$ implies that an electron cotunneled from electrode $\alpha$ to electrode $\beta$, changing the system from state $l$ to state then $k$; if the process is elastic $l=k$. Elastic cotunneling processes do not appear in Eq.\eqref{General master equation} as they do not affect the molecular probabilites $\mathbf{P}(t)$, while elastic cotunneling events involving the same electrode $\alpha=\beta$ are not included even in the $n$-resolved master equation as they neither change the jump number or the molecular state.

The cotunneling rates are calculated by going to fourth order in $H_{T}$ in Eq.\eqref{Generalised Fermi's golden rule} and Eq.\eqref{T matrix}. To avoid divergences in the T-matrix calculated cotunneling rates, and overcounting of sequential processes, one must use the regularisation procedure noted first by Averin \cite{Averin1994} and detailed separately by Koch $\text{\it{et al.}}$ \cite{Koch2006,Koch2004} and Turek and Matveev \cite{Turek2002}. The details and specific form of the lengthy cotunneling rates for an Anderson impurity are in Appendix \ref{Cotunneling rates}.

\subsection{Holstein model}

In the limit $U\rightarrow \infty$, the molecule cannot be doubly occupied and behaves as a single level coupled to a vibrational mode: the Holstein model, with the Hamiltonian

\begin{align}
H_{M} & = \varepsilon_{0} a^{\dagger}a + \lambda(b^{\dagger} + b)a^{\dagger}a+\omega b^{\dagger}b.
\end{align}

After the Lang-Firsov transformation \cite{Lang1963} the molecular Hamiltonian reduces to 
\begin{align}
H_{M} & = \varepsilon \tilde{a}^{\dagger}\tilde{a} + \omega \tilde{b}^{\dagger}\tilde{b},
\end{align}
with $\varepsilon = \varepsilon_{0} - \frac{\lambda^{2}}{\omega}$ and new fermionic $\tilde{a}^{\dagger}\tilde{a}$ and bosonic $\tilde{b}^{\dagger}\tilde{b}$ operators obtained from the original operators via the canonical transformation. The molecular eigenstate is now described by two quantum numbers $|nq\rangle$, where $n=\left\{0,1\right\}$ is the electron occupation number and $q = \left\{0,1,...,+\infty\right\}$ is the vibrational quanta occupation number. The associated eigenenergy is $E_{nq} = \varepsilon n + \omega q$. The Fourier transformed vector is then
\begin{align}
\mathbf{P}(\chi,t) & = \left[\begin{array}{c} P_{00}(\chi,t) \\ P_{10}(\chi,t) \\ P_{01}(\chi,t) \\ P_{11}(\chi,t) \\ . \\ . \\  . \\  P_{0N}(\chi,t) \\ P_{1N}(\chi,t) \end{array} \right],
\end{align}
which is described by the master equation \cite{Mitra2004}

\begin{figure*}
	\subfloat[\label{Fig1a_Cotunn_Current}]{\includegraphics[scale=0.5]{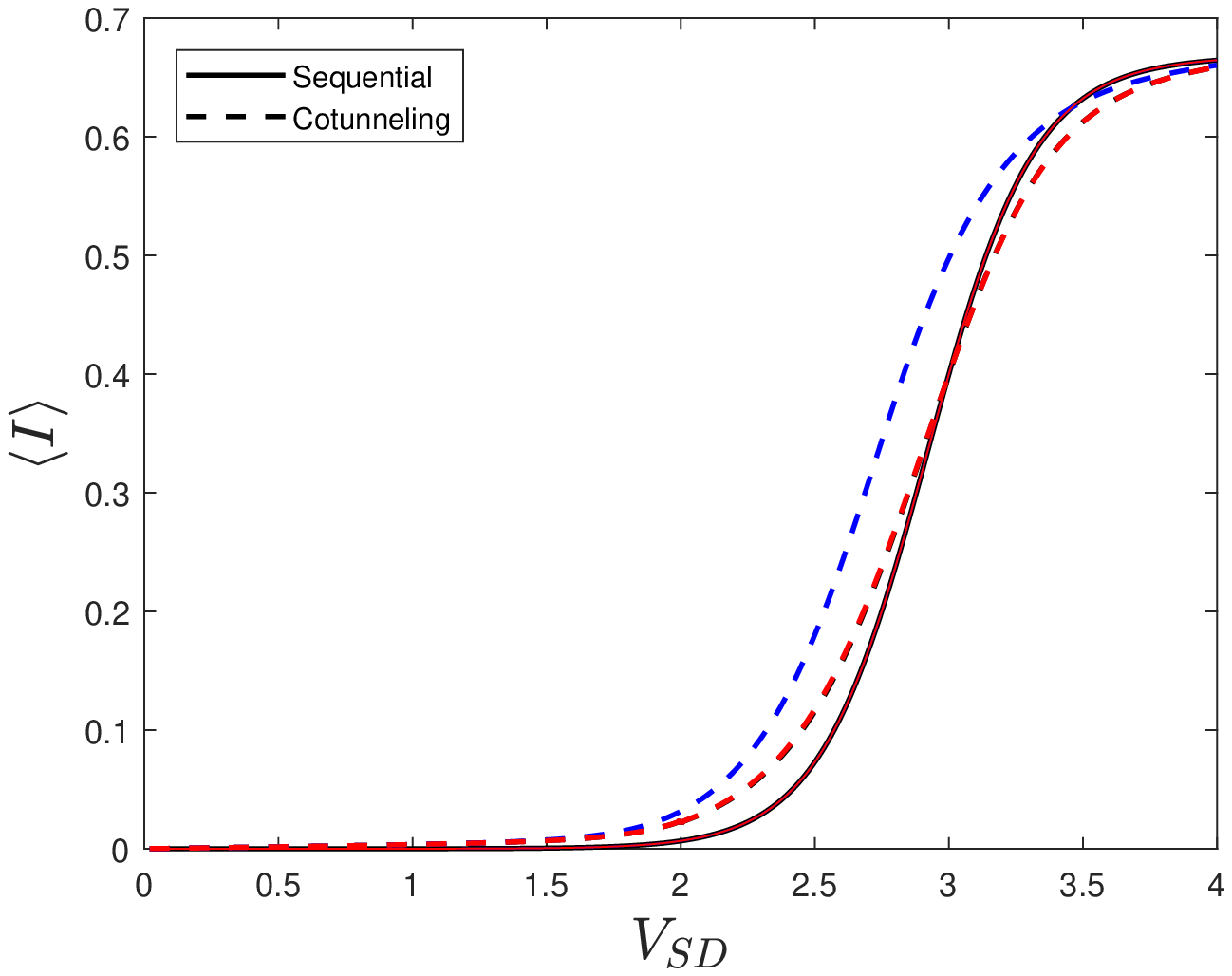}}
	\subfloat[\label{Fig1b_Cotunn_Fano}]{\includegraphics[scale=0.5]{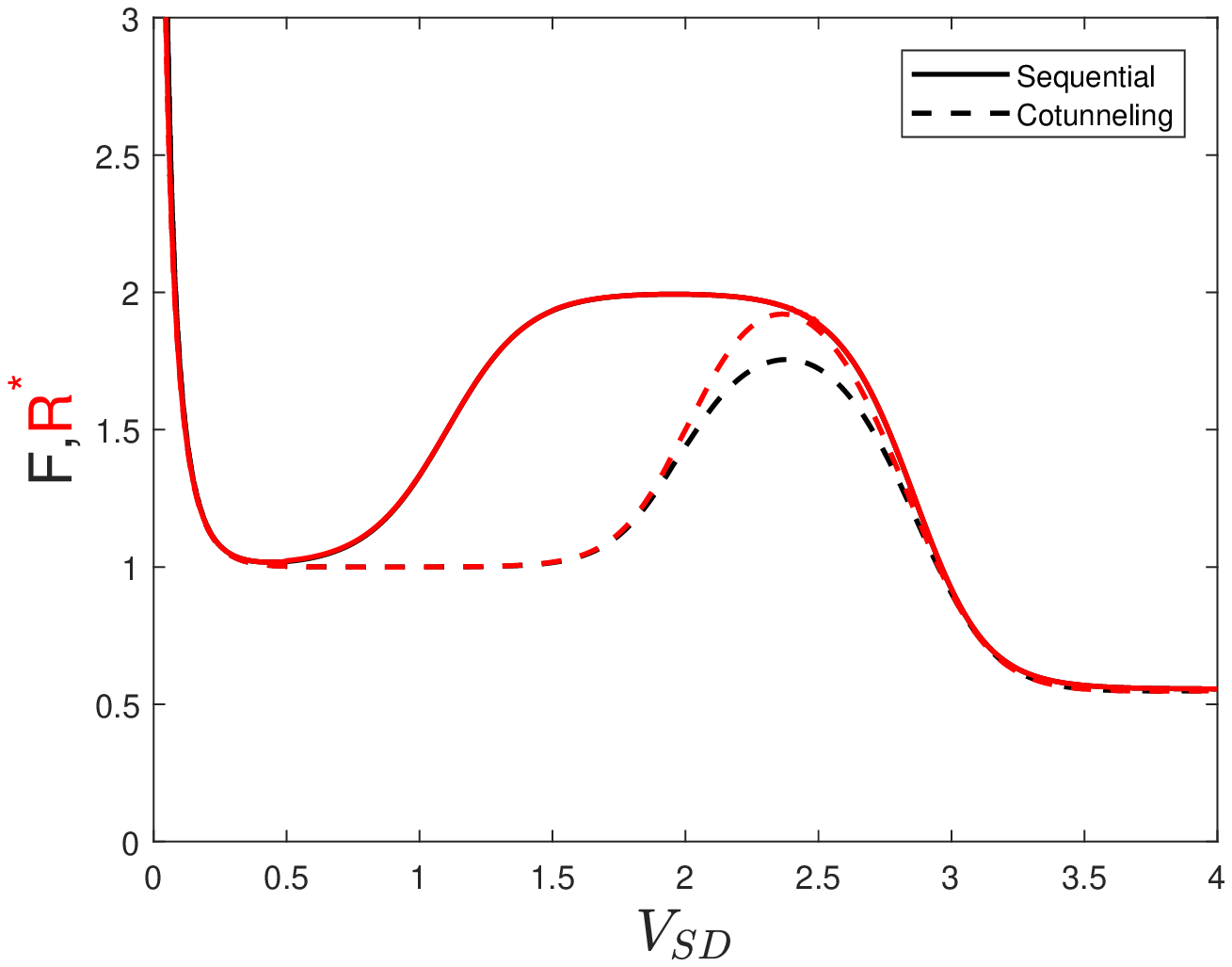}}
	\caption{Colour online. (a) The exact sequential and cotunneling current $\langle I \rangle = \left(\mathbf{I},(\mathbf{J}_{F} - \mathbf{J}_{B})\bar{\mathbf{P}}\right)$ (black) compared to the respective predictions from the FPTD $\frac{1}{\langle\tau\rangle^{*}}$ (red) and WTD $\frac{1}{\langle\tau\rangle_{F}} - \frac{1}{\langle\tau\rangle_{B}}$ (blue). Current is measured in units of $\frac{\gamma}{2}$ and voltage is measured in $\text{meV}$. In (b) the randomness parameter $R^{*}$ calculated from Eq.\eqref{RP FPTD definition} is compared to the Fano factor $F$ defined in Eq.\eqref{Fano factor definition}. The energies of the spin-split electronic levels are $\varepsilon_{\uparrow}=0.5\text{meV}$ and $\varepsilon_{\downarrow}=-1.5\text{meV}$, the Coulomb repulsion is $U=4\text{meV}$, $T=75\mu\text{eV}$, and $\gamma=0.5 T$. The source and drain Fermi energies are symmetric about zero: $\mu_{S} = -\mu_{D} = V_{SD}/2$.}
\end{figure*}

\begin{widetext}
\begin{align}
\dot{P}_{0q}(\chi,t) & = \sum_{q'} \left(\Gamma^{S}_{0q;1q'} + \Gamma^{D}_{0q;1q'}e^{i\chi}\right)P_{1q'}(\chi,t) - \sum_{\alpha q'} \Gamma^{\alpha}_{1q';0q}P_{0q}(\chi,t), \label{P0 rate equation holstein} \\ 
\dot{P}_{1q}(\chi,t) & = \sum_{q'} \left(\Gamma^{S}_{1q;0q'} + \Gamma^{D}_{1q;0q'}e^{-i\chi}\right)P_{0q'}(\chi,t) - \sum_{\alpha q'} \Gamma^{\alpha}_{0q';1q}P_{1q}(\chi,t) \label{P1 rate equation holstein}.
\end{align}
\end{widetext}
Evidently, there can be any number of interacting vibrations and $\mathbf{P}(\chi,t)$ is a vector of infinite length. In practice, however, we only include a finite number $N$ of vibrations in the transport calculations, so that $\mathbf{P}(\chi,t)$ and $\mathbf{I}$ are vectors of length $2(N+1)$. Likewise, the Liouvillian and two jump operators are $2(N+1)\times2(N+1)$ matrices. $\mathbf{J}_{F}$ and $\mathbf{J}_{B}$ are explicitly written in Appendix \ref{Vibration jump operators}.

The rates are 
\begin{align}
\Gamma^{\alpha}_{1q';0q} & = \gamma\left|X_{q'q}\right|^{2}n_{F}(\varepsilon + \omega(q'-q) - \mu_{\alpha}) \text{ and } \\
\Gamma^{\alpha}_{0q';1q} & = \gamma\left|X_{q'q}\right|^{2}\left[1-n_{F}(\varepsilon + \omega(q'-q) - \mu_{\alpha})\right],
\end{align}

where the Franck-Condon matrix elements are 
\begin{align}
X_{qq'} & = \langle q | e^{-\lambda(b^{\dagger}-b)}|q'\rangle.
\end{align}

We consider two distinct vibrational scenarios. So far we have described transport when the phonons are unequilibrated, but when the phonons are in equilibrium with an external bath at temperature $T_{V}$, the rate equation reduces to that of a single resonant level, with Liouvillian

\begin{align}
\mathbf{L}_{0} & = \left[\begin{array}{cc} -T_{10} & T_{01}^{S} \\ T^{S}_{10} & -T_{01} \end{array}\right],
\end{align} and jump operators

\begin{widetext}
\begin{align}
\mathbf{J}_{F} & = \left[\begin{array}{cc} 0 & T_{01}^{D} \\ 0 & 0 \end{array}\right], & \mathbf{J}_{B} & = \left[\begin{array}{cc} 0 & 0 \\ T_{10}^{D} & 0 \end{array}\right],
\end{align}
\end{widetext}
where the probabilities are defined with the ansatz 
\begin{align}
P_{nq}(\chi,t) & = P_{n}(\chi,t)\frac{e^{-q\omega/T_{V}}}{1-e^{-\omega/T_{V}}};
\end{align}
the total transition rates are $T_{kl} = \sum\limits_{\alpha} T_{kl}^{\alpha}$; and the individual transition rates are 
\begin{align}
T^{\alpha}_{kl} = \sum_{qq'} \Gamma^{\alpha}_{kq;lq'} \frac{e^{-q\omega/T_{V}}}{1-e^{-\omega/T_{V}}}.
\end{align}


\section{Results} \label{Results}

In this section we compare the full counting statistics, waiting time distribution, and first-passage time distribution for the three transport scenarios.

\subsection{Sequential tunneling and cotunneling through an Anderson impurity}

In a previous paper, we examined, via waiting times, the effect cotunneling has on strong electron-electron interactions in an Anderson impurity \cite{Rudge2018}. In order to include elastic cotunneling events, it is necessary to use the $n$-resolved master equation to calculate the waiting times. Unfortunately, when bidirectional transitions are included the definition of the idle-time probability in Eq.\eqref{Idle time probability} is no longer satisfied \cite{Thomas2013}. One may of course treat the transport as unidirectional and calculate the waiting times in either direction; Fig.(\ref{Fig1a_Cotunn_Current}) shows that doing this, in a regime where backtunneling processes are present, does not reproduce the correct cotunneling current from $\frac{1}{\langle\tau\rangle_{F,co}} - \frac{1}{\langle\tau\rangle_{B,co}}$. The first-passage time prediction $\frac{1}{\langle\tau\rangle_{co}^{*}}$, on the other hand, reproduces the cotunneling current; the black and red dashed lines in Fig.(\ref{Fig1a_Cotunn_Current}) coincide. Over the voltage range $2-3\text{meV}$ there is a small numeric difference between $\langle I \rangle_{co}$ and $\frac{1}{\langle\tau\rangle_{co}^{*}}$, since the transport is non-renewal, but it is not visible to the naked eye.  In comparison, the sequential current is exactly reproduced by $\frac{1}{\langle\tau\rangle_{F,seq}} - \frac{1}{\langle\tau\rangle_{B,seq}}$, since sequential tunneling is described completely by a standard master equation, and $\frac{1}{\langle\tau\rangle_{seq}^{*}}$, since the sequential transport is always renewal. One can see this in Fig.(\ref{Fig1a_Cotunn_Current}), as all solid lines coincide.

\begin{figure}
{\includegraphics[scale=0.5]{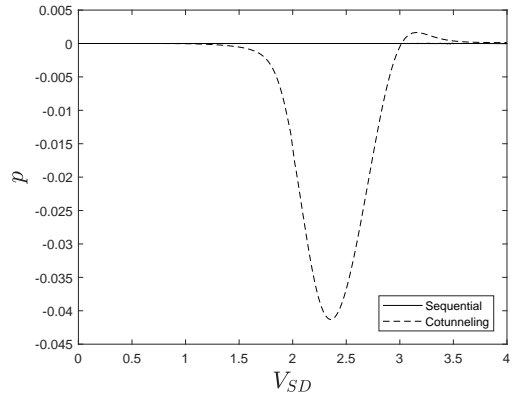}}
\caption{The Pearson correlation coefficient for sequential tunneling and cotunneling, defined in Eq.\eqref{pearson FPTD definition}, calculated from the FPTD only. The energies of the spin split electronic levels are $\varepsilon_{\uparrow}=0.5\text{meV}$ and $\varepsilon_{\downarrow}=-1.5\text{meV}$, the Coulomb repulsion is $U=4\text{meV}$, $T=75\mu\mbox{eV}$, and $\gamma=0.5 T$.}
\label{Fig2_Cotunn_pearson}
\end{figure}

These difficulties previously kept us to the large bias voltage regime. It is well known in the literature, however, that in a certain voltage range inelastic cotunneling processes cause telegraphic switching in an Anderson impurity, which is now accessible to fluctuating time statistics via the FPTD. The Fano factor in Fig.(\ref{Fig1b_Cotunn_Fano}) is a reproduction of Fig.(3) from Ref.[\onlinecite{Thielmann2005a}], with the first-passage and waiting time randomness parameters included as well. The molecule is in the Coulomb blockade regime $U\gg V_{SD}$, so that transport through one $\sigma$ level blocks transport through the other $\bar{\sigma}$ level. At small voltages, for sequential tunneling, Belzig \cite{Belzig2005} has shown that the molecule goes through cycles of transferring $n-1$ electrons quickly through the $\uparrow$ level until it gets ``stuck'' in the $\downarrow$ level. The cycle completes when the $\downarrow$ empties via thermal broadening of the drain occupation function. The noise is thus characteristic of a sum of Poissonian processes, each transferring $n$ electrons, and weighted by $(\frac{1}{2})^{n}$. In contrast, when cotunneling processes are included, super-Poissonian noise arises from a different mechanism. Note that from here we have not included cumulants from the forward cotunneling WTD, such as $R$ and $p$, since in this voltage regime the transport is bidirectional. 

At low bias $V_{SD}/2 < \varepsilon_{\uparrow}$ the only transport mechanism is elastic cotunneling through the empty $\uparrow$ level or the occupied $\downarrow$ level; the noise is Poissonian and $F=1$. When $\varepsilon_{\downarrow} > V_{SD}/2 \geq \varepsilon_{\uparrow}$ the $\uparrow$ level may be occupied via sequential tunneling, and there are now two main transport channels: elastic cotunneling through the $\downarrow$ level and sequential tunneling through $\uparrow$ level. Telegraphic switching between these two processes occurs when inelastic cotunneling induces a spin flip $\sigma\rightarrow\bar{\sigma}$ or the molecule relaxes from $\uparrow\rightarrow\downarrow$ via sequential tunneling. Super-Poissonian noise is a symptom of the switching between two Poissonian processes, which is the main observation of Fig.(3) from Ref.[\onlinecite{Thielmann2005a}] and is evident in Fig.(\ref{Fig1b_Cotunn_Fano}). In this voltage regime the transport is evidently non-renewal as $R^{*}$ does not match the Fano factor; there should thus be correlations between subsequent first-passage times. 

Positive correlations in the waiting times are known to occur for systems undergoing telegraphic switchin: for example, in systems with spin-polarised electrodes \cite{Ptaszynski2017}. The current switches between two transport channels, each with a characteristic waiting time. When in the faster channel, waiting times between successive tunnelings are more likely to be shorter, and vice versa in the slower channel. Consequently, waiting times between successive tunnelings to the drain are positively correlated; one would expect that there would be similar positive correlations between the first-passage times. In Fig.(\ref{Fig2_Cotunn_pearson}), however, we report intriguing negative correlations between successive first-passage times. That is, if the jump number reaches $+1$ after a short time-interval, it is more likely that the time until the jump number next reaches $+1$ will be long. The values are small enough, however, that the first-passage times are still not considered correlated. A possible explanation is that the time it takes for the transport to switch between the two mechanisms is comparable to the average first-passage times for the two channels, so that the system does not spend long enough in either channel for subsequent first-passage times to become positively correlated.

\begin{figure*}
	\subfloat[\label{Fig3a_Vib_Current}]{\includegraphics[scale=0.5]{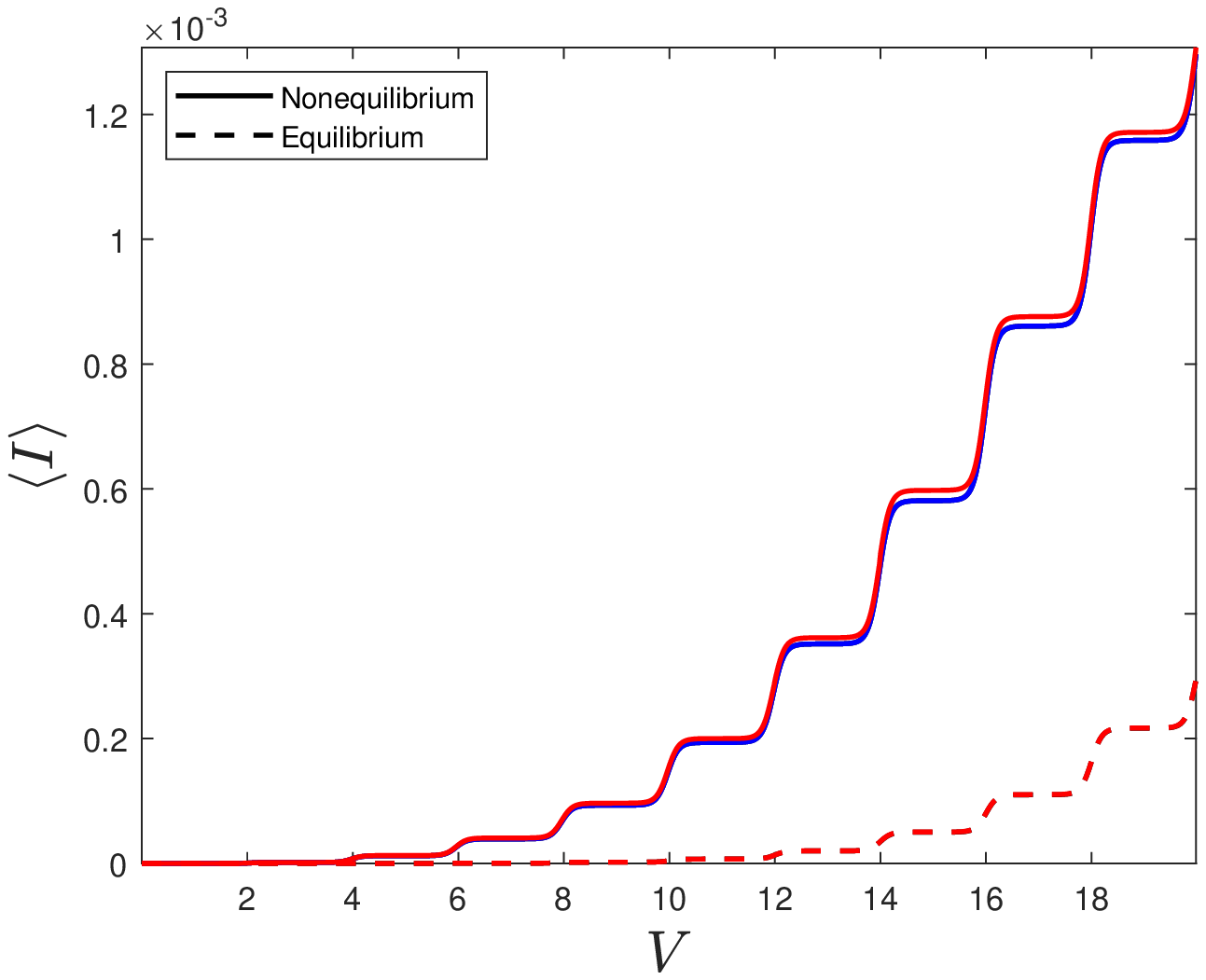}}
	\subfloat[\label{Fig3b_Vib_Fano}]{\includegraphics[scale=0.5]{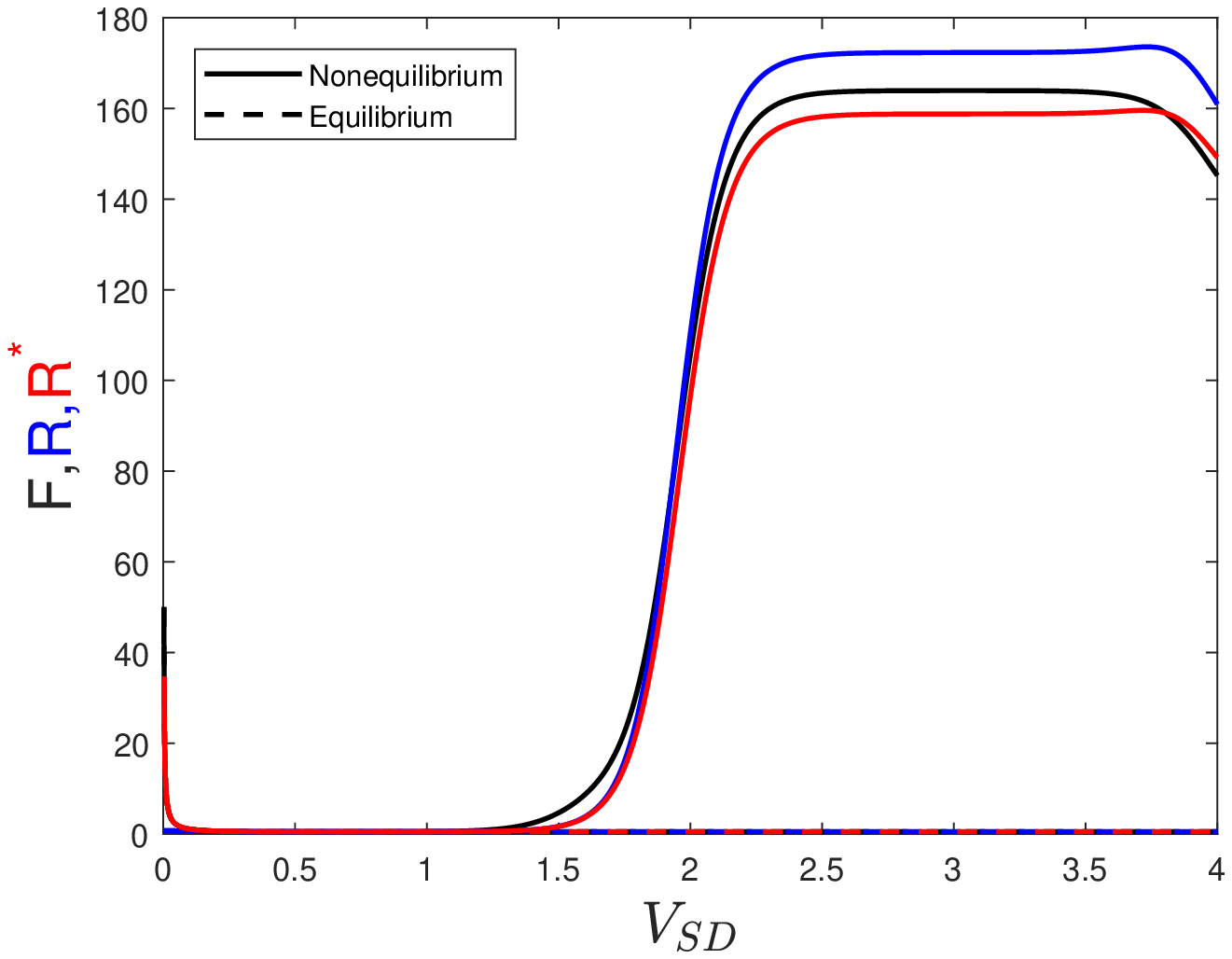}}
    	\caption{Colour online.  (a) The exact current for equilibrium and non-equilibrium phonons $\langle I \rangle = \left(\mathbf{I},(\mathbf{J}_{F} - \mathbf{J}_{B})\bar{\mathbf{P}}\right)$ (black) compared to the respective predictions from the FPTD $\frac{1}{\langle\tau\rangle^{*}}$ (red) and WTD $\frac{1}{\langle\tau\rangle_{F}} - \frac{1}{\langle\tau\rangle_{B}}$ (blue). In (b) the randomness parameters $R^{*}$ calculated from Eq.\eqref{RP FPTD definition} and $R$ calculated from Eq.\eqref{RP WTD} are compared to the Fano factor $F$ defined in Eq.\eqref{Fano factor definition}. Note that here the WTD current prediction is $\frac{1}{\langle\tau\rangle_{F}}-\frac{1}{\langle\tau\rangle_{B}}$, but $R$ is calculated only from the forward WTD. The vibrationally adjusted energy level is $\varepsilon = 0$, the vibrational frequency is $\omega = 1$, $T_{S} = T_{D} = 0.05$, the electron-phonon coupling strength is $\lambda = 4$, $\gamma_{\alpha} = \frac{\gamma}{2} = 0.01$, and the equilibrium phonons are kept at a vibrational temperature of $T_{V} = 0.05$. The source and drain chemical potentials are shifted symmetrically about zero: $\mu_{S} = -\mu_{D} = V_{SD}/2$. We use units of $\omega$ for $\langle I \rangle$ ($e\omega$ if we reintroduce $\hbar\text{ and }e$) and units of $\omega$ for all energy parameters as well (or $\hbar\omega/e$).}
\end{figure*}

\begin{figure}
{\includegraphics[scale=0.5]{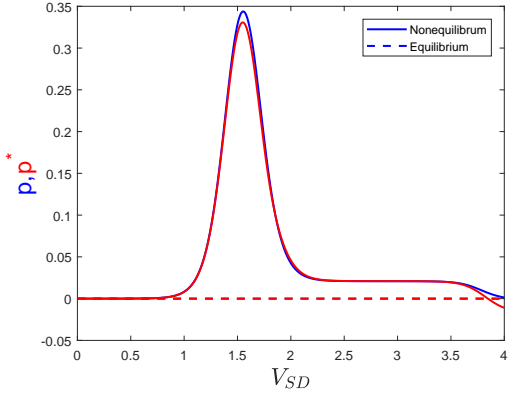}}
\caption{Colour online. The Pearson correlation coefficient for equilibrium and nonequilibrium vibrations, defined in Eq.\eqref{pearson WTD definition} and Eq.\eqref{pearson FPTD definition}, calculated from the WTD and the FPTD, respectively. The vibrationally adjusted energy level is $\varepsilon = 0$, the vibrational frequency is $\omega = 1$, $T_{S} = T_{D} = 0.5$, the electron-phonon coupling strength is $\lambda = 4$, $\gamma^{\alpha} = \frac{\gamma}{2} = 0.01$, and the equilibrium phonons are kept at a vibrational temperature of $T_{V} = 0.05$. Again, the electrode chemical potentials are shifted symmetrically about zero $\mu_{S} = -\mu_{D} = V_{SD}/2$, and the energy units are in $\omega$.}
\label{Fig4_Vib_pearson}
\end{figure}

\subsection{Vibrations with a finite bias voltage}

One of the authors (Kosov) has previously reported positive correlations in successive waiting times for transport through the Holstein model \cite{Kosov2017b}, which we now compare to first-passage times.

We consider transport through the Holstein model with fixed temperature and over a range of $V_{SD}$. In such a scenario, Koch and von Oppen have previously explained the different behaviour of equilibrated and unequilibrated current at low voltages \cite{Koch2005}, which originates from $X_{qq'}$ being suppressed on diagonal and close-to-diagonal elements under a strong electron-phonon coupling $\lambda \gg 1$. At small $T$ and $\text{eV}$, when the electrons are not energetic enough to reach highly excited phonon states, the current is thus suppressed, which is evident more for equilibrated phonons than out-of-equilibrium phonons, as multiple unequilibrated transitions can produce the required excited phonon states. At higher voltages, the current also displays the characteristic steplike structure, with steps occurring when the voltage is a multiple of $2$, since $\omega = 1$.

The total electronic current is correctly predicted from the WTD in Eq.\eqref{WTD Current definition} for both equilibrated and unequilibrated phonons; the blue and black dashed and solid lines in Fig.(\ref{Fig3a_Vib_Current}) exactly overlap. In comparison, $\frac{1}{\langle \tau \rangle^{*}}$ correctly predicts the current over all voltages for equilibrated but not unequilibrated phonons, implying that at the voltages where they differ the transport is non-renewal. One can see that the red dashed line in Fig.(\ref{Fig3a_Vib_Current}) coincides with both the blue and black dashed lines, but that there is a difference between the red solid line and the blue and black solid lines. Unlike the current plots for the other transport scenarios we consider, the discrepancy between the unequilibrated $\frac{1}{\langle\tau\rangle^{*}}\text{ and }\langle I \rangle$ due to renewal processes is large enough to be visually recognisable. As Fig.(\ref{Fig4_Vib_pearson}) shows, however, the largest correlations occur at voltages where the current, and thus the difference between $\frac{1}{\langle \tau \rangle^{*}}\text{ and }\langle I \rangle$, is negligible; consequently, we turn to the randomness parameters to analyse the non-renewal behaviour.

As found in Ref.[\onlinecite{Kosov2017b}], when phonons are unequilibrated, there exist strong correlations between successive waiting times, due to an elastic ``shortcut'' channel that opens in a small voltage range. In Fig.(\ref{Fig4_Vib_pearson}) we report similar correlations in the first-passage times over the same voltage range, albeit with a smaller peak. This is expected; due to backtunneling events the first-passage time can never be smaller than the waiting time $\tau^{*} \geq \tau$, and so correlations between first-passage times must be smaller. The Fano factor and randomness parameters in Fig.(\ref{Fig3b_Vib_Fano}) support this non-renewal behaviour. At low voltages $R$ fails to capture the divergence of the Fano factor and $R^{*}$. Between the $1\rightarrow2\text{eV}$ voltage range $F$ and $R^{*}$ differ, correctly predicting the non-renewal behaviour in Fig.(\ref{Fig4_Vib_pearson}). In comparison, $R$ actually crosses the $F$ at $\sim 1.9\text{eV}$, implying renewal behaviour, while at the same voltage the waiting time Pearson coefficient is still a non-zero $p \approx 0.05$. The correlations disappear when the phonons are forced to relax to equilibrium immediately, as the excited phonon state providing the elastic shortcut channel is no longer occupied. 


\begin{figure*}
	\subfloat[]{\includegraphics[scale=0.5]{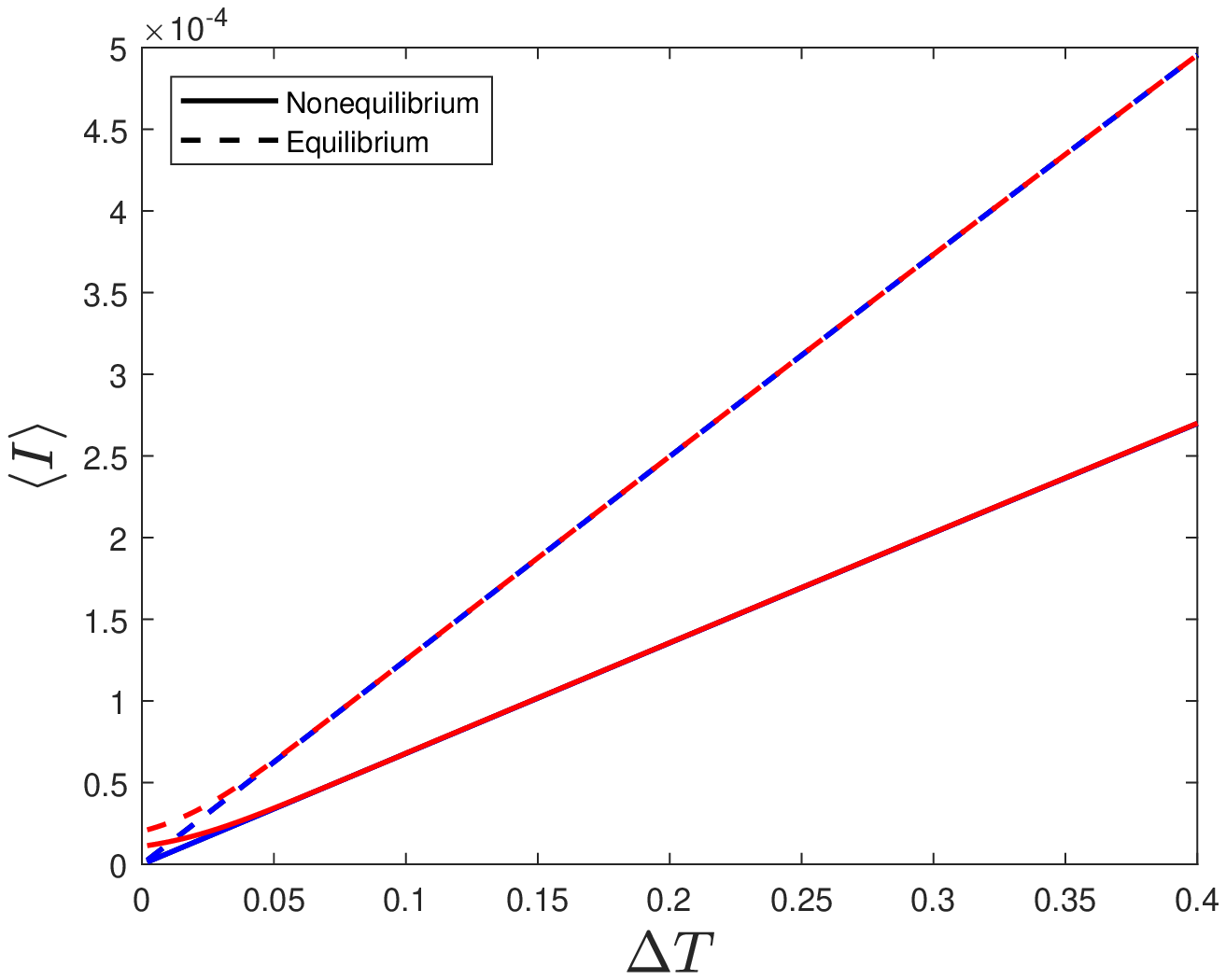}\label{Fig5a_TempGrad_Current}}
	\subfloat[]{\includegraphics[scale=0.5]{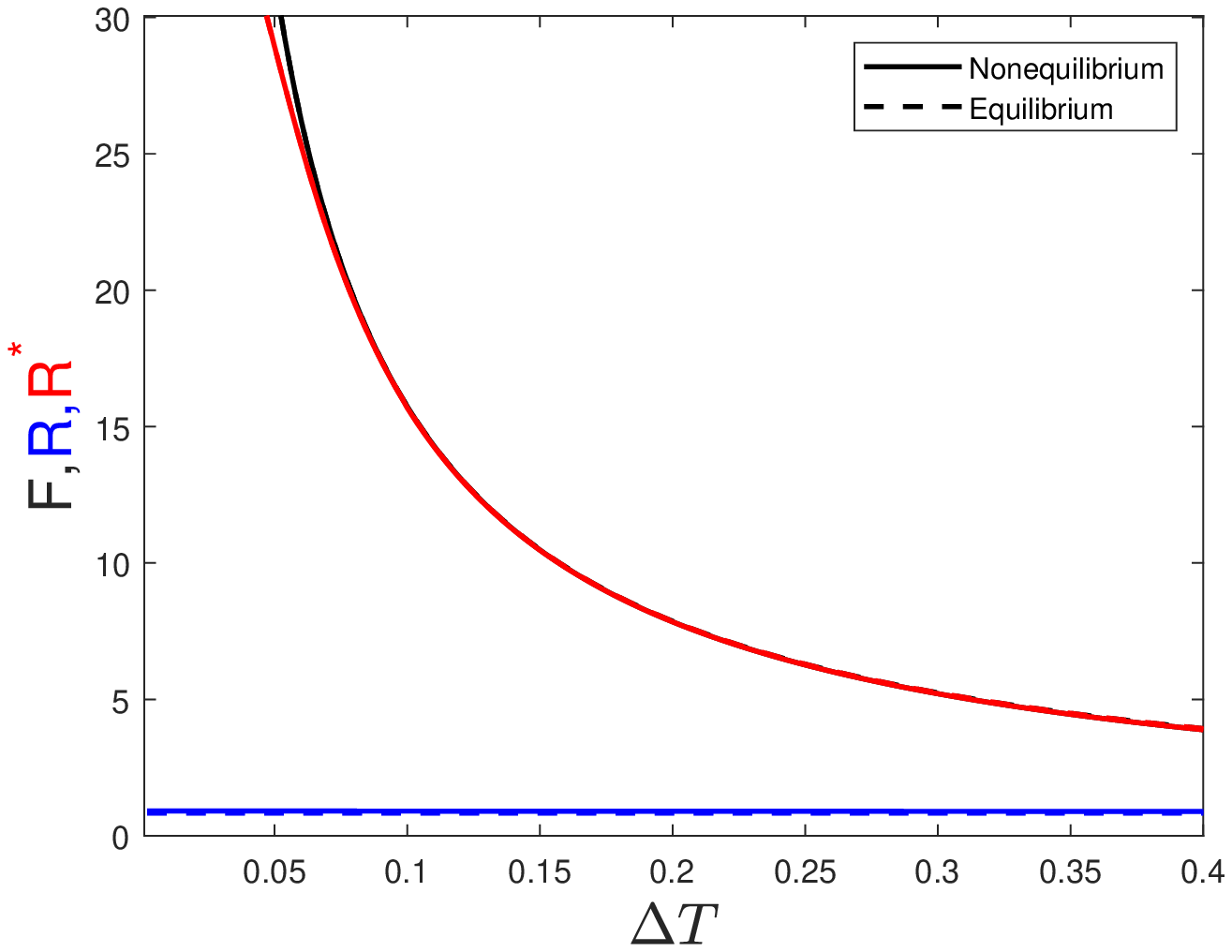}\label{Fig5b_TempGrad_Fano}}
    	\caption{Colour online.  Current (a) and Fano factor (b) predicted from the WTD (blue) and FPTD (red) and compared to the exact results (black) for equilibrium and nonequilibrium vibrations, as a function of the temperature gradient $\Delta T$. Again, the WTD prediction is $\frac{1}{\langle\tau\rangle_{F}}-\frac{1}{\langle\tau\rangle_{B}}$, but $R$ is calculated only from the forward WTD. The vibrationally adjusted energy level is $\varepsilon = 1$, the vibrational frequency is $\omega = 1$, the electron-phonon coupling strength is $\lambda = 1$, the voltage is $V_{SD} = 0$, $\gamma^{\alpha} = \frac{\gamma}{2} = 0.01$. The average temperature across the molecule is $\bar{T} = \frac{T_{S}+T_{D}}{2} = 0.75$ and the electrode temperatures are symmetric around $\bar{T}$: $T_{S} = \bar{T}+\Delta T/2$ and $T_{D} = \bar{T}-\Delta T/2$. The equilibrium phonons are kept at a vibrational temperature of $T_{V} = \bar{T}$. Again, energy and current units are in $\omega$.}
\end{figure*}

\subsection{Vibrations with a finite temperature gradient} \label{Temperature gradient results}

At finite temperatures, and when the electrode Fermi energies are approximately in resonance with $\varepsilon + \omega q$, the bidirectional transport is dominated by thermal noise. A recent experimental paper has shown, however, that electronic noise distinct from the thermal noise can be generated from a temperature gradient $\Delta T$ in the absence of a bias voltage, which they label delta-T noise \cite{Lumbroso2018}. We finish our investigation by exploring the effect a temperature gradient has on the first-passage times for sequential tunneling through the Holstein model.

Since the transport in this scenario is described by a standard master equation, the exact current matches the prediction from the WTD in Eq.\eqref{WTD Current definition}, which is shown in Fig.(\ref{Fig5a_TempGrad_Current}); all black and blue lines coincide. The FPTD current prediction from Eq.\eqref{FPTD Current definition}, on the other hand, differs from $\langle I \rangle$ at low $\Delta T$, implying non-renewal behaviour in this regime; otherwise, the red FPTD lines coincide with all blue and black lines. Interestingly, the equilibrium current is larger than the non-equilibrium current, which is understood by examining $X_{qq'}$ and the phonon occupation probabilities. For $\lambda = 1$ and $\omega = 1$ the Franck-Condon factor is largest for elastic $q=0$ transitions. These transitions are more accessible to phonons in equilibrium, as the $q=0$ state has a larger relative occupation probability for equilibrium phonons than non-equilibrium phonons: 

\begin{align}
\frac{P^{eq}_{q=0}}{P^{eq}_{q=1}} & > \frac{P^{ne}_{0;q=0}+P^{ne}_{1;q=0}}{P^{ne}_{0;q=1}+P^{ne}_{1;q=1}}. 
\end{align}
We note that for large $\Delta T$ the total current is the same order as that induced by a voltage bias in Fig.(\ref{Fig3a_Vib_Current}).

In Fig.(\ref{Fig5b_TempGrad_Fano}) $F$ diverges as $\Delta T \rightarrow 0$, which is expected since thermal processes dominate in this regime. At larger $\Delta T$ the noise no longer diverges but is still super-Poissonian, and distinct from the shot noise generated by a voltage bias. We attribute this to the relatively large temperatures; even the smallest drain temperature we consider is large enough to cause backtunnelings from the drain, hence the super-Poissonian partition noise normally associated with a voltage bias.

It is evident that $R^{*} \neq F$ at small $\Delta T$, implying that the transport is non-renewal in this temperature range. Despite this non-renewal behaviour, we have chosen not to display the Pearson correlations: they are non-zero but small. At larger $\Delta T$ when the renewal assumption is satisfied, $R^{*}$ successfully predicts $F$, while $R$ clearly does not as it only includes unidirectional transitions.

\section{Conclusion} \label{Conclusion}

In this paper we have compared the waiting time statistics to the first-passage time statistics for a molecule with strong electron-electron or electron-phonon interactions. All methods rely on a form of the master equation, in which the on-site interactions are treated exactly and the coupling between the molecule and electrode is treated perturbatively under the Born-Markov approximation. In bidirectional transport scenarios we have demonstrated the effectiveness of using first-passage times as opposed to waiting times. In particular, we compared the total current to the current computed from the WTD and FPTD, the Fano factor to the WTD and FPTD randomness parameters, and the correlation between subsequent waiting and first-passage times. We first examine sequential tunneling and cotunneling through an Anderson impurity. When a full transport description requires use of the $n$-resolved master equation, the WTD can be calculated in the forward tunneling regime only. Such difficulties arise when including elastic cotunneling, but do not apply to first-passage times; we demonstrate that there are unexpected, minor negative correlations between the first-passage times even though the transport displays telegraphic switching. Next, we analysed the Holstein model at finite bias voltage, in which correlations between successive first-passage times are smaller than correlations between successive waiting times when backtunneling events are present. Finally, we considered transport through the Holstein model at zero bias voltage and a finite temperature gradient. The temperature gradient induces a current, but also significant backtunneling events. Hence, the forward waiting time $\tau_{F}$ was incapable of accurately predicting the Fano factor, as opposed to the first-passage time $\tau^{*}$.

\appendix \label{appendix}

\section{Cotunneling rates} \label{Cotunneling rates}

Cotunneling rates calculated from the T-matrix approach are regularised according to the procedure outlined by Koch et al \cite{Koch2006,Koch2004}. Calculating cotunneling rates for transport through an Anderson impurity, in particular, has been thoroughly covered by in Ref.[\onlinecite{Kaasbjerg2015,Rudge2018}], to which we direct the reader for a comprehensive overview.

The elastic cotunneling rates are 

\begin{widetext}
\begin{align}
\Gamma_{nn}^{\alpha\beta} = \frac{\gamma^{2}}{2\pi}\lim_{\eta\rightarrow0^{+}}\int d\varepsilon \Big| \frac{1}{\varepsilon-E_{1}+i\eta} \pm \frac{1}{\varepsilon-E_{2}+i\eta}\Big| n_{F}(\varepsilon-\mu_{\alpha})[1-n_{F}(\varepsilon-\mu_{\beta})], \label{Elastic cotunneling rate definition}
\end{align}
\end{widetext}
where $E_{1}\text{ and }E_{2}$ are derived from the cotunneling pathways involved in the rate, and the $\pm$ is only positive for elastic tunneling through an initially empty or initially doubly occupied system. Similarly, inelastic cotunneling rates have the general form 
\begin{widetext}
\begin{align}
\Gamma_{\bar{\sigma}\sigma}^{\alpha\beta} = \frac{\gamma^{2}}{2\pi}\lim_{\eta\rightarrow0^{+}}\int d\varepsilon \Big| \frac{1}{\varepsilon-\varepsilon_{\bar{\sigma}}-U+i\eta} - \frac{1}{\varepsilon-\varepsilon_{\bar{\sigma}}-i\eta}\Big| n_{F}(\varepsilon-\mu_{\alpha})[1-n_{F}(\varepsilon-\mu_{\beta}+\varepsilon_{\sigma}-\varepsilon_{\bar{\sigma}})], \label{Inelastic cotunneling rate definition}
\end{align}
\end{widetext}
where the broadening of the intermediate virtual state $i\eta$ in Eq.\eqref{Elastic cotunneling rate definition} and Eq.\eqref{Inelastic cotunneling rate definition}was first noted by Averin. 

The rates are then split into a sum via:

\begin{widetext}
\begin{align} 
\left| \frac{1}{\varepsilon-E_{1}+i\eta} \pm \frac{1}{\varepsilon-E_{2}-i\eta} \right| & = \frac{1}{(\varepsilon-E_{1})^{2}+\eta^{2}}+\frac{1}{(\varepsilon-E_{2})^{2}+\eta^{2}}\pm2\Re\left\{\frac{1}{\varepsilon-E_{1}+i\eta}\cdot\frac{1}{\varepsilon-E_{2}-i\eta}\right\}.
\end{align}
\end{widetext}

All that remains is to compute the two types of integrals:
\begin{widetext}
\begin{align}
\lim_{\eta\rightarrow 0}\Re\int d\varepsilon \frac{n_{F}(\varepsilon - \mu_{\alpha})\left[1-n_{F}(\varepsilon-\mu_{\beta})\right]}{(\varepsilon - E_{1}-i\eta)\cdot(\varepsilon - E_{2}+i\eta)} & = 
\frac{n_{B}(\mu_{\beta}-\mu_{\alpha})}{E_{2}-E_{1}}\Re\Big\{\psi^{(1)}\left(E^{\alpha}_{-,2}\right) - \psi^{(1)}\left(E^{\alpha}_{+,1}\right) \\
& \quad\quad\quad\quad\quad\quad\quad\quad - \psi^{(1)}\left(E^{\beta}_{-,2}\right) + \psi^{(1)}\left(E^{\beta}_{+,1}\right)\Big\}, \\
\lim_{\eta\rightarrow 0}\int d\varepsilon \frac{n_{F}(\varepsilon - \mu_{\alpha})\left[1-n_{F}(\varepsilon-\mu_{\beta})\right]}{(\varepsilon - E_{k})^{2}+\eta^{2}} - \mathcal{O}(\eta^{-1}) & = 
\frac{n_{B}(\mu_{\beta}-\mu_{\alpha})}{2\pi T}\Im\left\{\psi^{(1)}\left( E^{\beta}_{+,k}\right) - \psi^{(1)}\left( E^{\alpha}_{+,k}\right)\right\}.
\end{align}
\end{widetext}

Here, $E^{\alpha}_{\pm,k} = 1/2\:\pm \: i\frac{\mu_{\alpha} - E_{k}}{2\pi T}$; $\psi(x)$ and $\psi^{(1)}(x)$ are the digamma and trigamma functions, respectively; and the Bose-Einstein distribution is 

\begin{align}
n_{B}(\mu_{\beta}-\mu_{\alpha}) & = \frac{1}{e^{(\mu_{\beta}-\mu_{\alpha})/T}-1}.
\end{align} 

\section{Jump operators for the Holstein model} \label{Vibration jump operators}

The jump operators defined from the master equation in Eq.\eqref{P0 rate equation holstein} and Eq.\eqref{P1 rate equation holstein} are 

\begin{widetext}
\begin{align}
\mathbf{J}_{F} & = \left[\begin{array}{ccccccccc}
0 & \Gamma^{D}_{00;10} & 0 & \Gamma^{D}_{00;11} & 0 & \cdots & \cdots & \cdots & \Gamma^{D}_{00;1N} \\ 
0 & 0 & 0 & 0 & 0 & & & & 0 \\
0 & \Gamma^{D}_{01;10} & 0 & \Gamma^{D}_{01;11} & 0 & & & & \Gamma^{D}_{01;1N} \\
\vdots & & & \ddots & & & & & \vdots \\ 
\vdots &  & & & \ddots & & & & \vdots \\ 
\vdots &  & & & & \ddots & & & \vdots \\ 
0 & & & & & 0 & 0 & 0 & 0 \\
0 & & & & & 0 & \Gamma^{D}_{0N;1(N-1)} & 0 & \Gamma^{D}_{0N;1N} \\
0 & \cdots & \cdots & \cdots & \cdots & 0 & 0 & 0 & 0 \\
\end{array}\right], \text{ and } \\
\mathbf{J}_{B} & = \left[\begin{array}{ccccccccc}
0 & 0 & 0 & 0 & \cdots & \cdots & \cdots & 0 & 0 \\
\Gamma^{D}_{10;00} & 0 & \Gamma^{D}_{10;01} & 0 & & & & \Gamma^{D}_{10;0N} & 0 \\ 
0 & 0 & 0 & 0 & & & & 0 & 0 \\
\Gamma^{D}_{11;00} & 0 & \Gamma^{D}_{11;01} & 0 & & & & \Gamma^{D}_{11;0N} & 0 \\
\vdots & & & & \ddots & & & & \vdots \\ 
\vdots & & & & & \ddots & & & \vdots \\ 
\vdots & & & & & & \ddots & & \vdots \\ 
0 & & & & & 0 & 0& 0 & 0 \\
\Gamma^{D}_{1N;00} & \cdots & \cdots & \cdots & \cdots & \Gamma^{D}_{1N;0(N-1)} & 0 & \Gamma^{D}_{1N;0N} & 0 \\
\end{array}\right],
\end{align}
\end{widetext}

\bibliography{Main_text_incl._figures}

\end{document}